\documentclass[letterpaper, 10 pt, conference]{ieeeconf}  

\IEEEoverridecommandlockouts                              %
\usepackage{graphicx}
\usepackage{hyperref}

\title{\LARGE \bf
An MPI-Based Python Framework for Distributed Training with Keras
}%

\author{Dustin Anderson$^{1}$, Jean-Roch Vlimant and Maria Spiropulu \\ California Institute of Technology, 1200 E.\ California Blvd, Pasadena, CA 91125
\thanks{Contact author dustin.james.anderson\@cern.ch}}%
\begin{document}

\maketitle

\begin{abstract}

We present a lightweight Python framework for distributed training of neural networks on multiple GPUs or CPUs. The framework is built on the popular Keras machine learning library. The Message Passing Interface (MPI) protocol is used to coordinate the training process, and the system is well suited for job submission at supercomputing sites.  We detail the software's features, describe its use, and demonstrate its performance on systems of varying sizes on a benchmark problem drawn from high-energy physics research.  

\end{abstract}


\section{Introduction}

Recent progress in machine learning has enabled deep neural networks (DNNs) to advance the state of the art in a wide range of problem domains, from computer vision to high energy physics~\cite{deeplearning}~\cite{hepdeeplearning}. As the applicability of DNNs has broadened, there have been efforts to develop user-friendly tools for building them.  Software packages such as Keras~\cite{keras} and TFLearn~\cite{tflearn} facilitate the construction and training of deep neural networks, offering a flexible interface for combining common model components and configuring the optimization process.  

Large model sizes and long training times have motivated the development of distributed training algorithms for DNNs~\cite{downpour}~\cite{easgd}.  These algorithms work by splitting the training task across multiple concurrent processes, which can be threads on a single machine or jobs spread across the nodes of a cluster. The speed-up provided by distributed algorithms is relevant when fast training is critical, such as when iterating on model choice during development, or when retraining a model on new data in a production environment.  

Despite the rise of convenient model-building software packages such as Keras, there are few tools for interfacing these packages with distributed training algorithms.  In this paper we introduce a lightweight Python framework, \textit{mpi\_learn}, that provides a straightforward means of training Keras models in a distributed fashion.  The framework is built on the Message Processing Interface (MPI) protocol~\cite{mpi} and can operate on personal machines, multi-GPU servers, and large supercomputing sites alike.  

\section{Related Work}

The package described here was written during the summer of 2016 and was motivated by the need for a mechanism to parallelize the training of models that took several days to converge.
It has been used for work for publications and conferences since early 2017.
This package, within the MPI framework, was developed concurrently with similar work on running distributed training of Keras models with Spark \cite{dist_keras}.

Since our experiments demonstrating the scaling of the algorithm, numerous articles have been produced studying theoretically and demonstrating experimentally the scaling of distributed training of deep neural network, targeting different training frameworks, including tensorflow \cite{horovod}.

The authors do not claim that their framework is better than any other framework.  This package was written for practical reasons 
in the observed absence of other tools fulfilling the same purpose.

\section{Package Overview}

The \textit{mpi\_learn} package is available on Github \cite{github}.  The prerequisites for using it are an \texttt{OpenMPI}~\cite{openmpi} installation and the \texttt{keras}~\cite{keras} and \texttt{mpi4py}~\cite{mpi4py} Python packages.  Support for the Theano~\cite{theano} and Tensorflow~\cite{tensorflow} backends to Keras is provided.  

\subsection{Training Algorithms}
The package supports two main distributed training algorithms based on stochastic gradient descent (SGD).  The default algorithm is Downpour SGD~\cite{downpour}, in which worker processes compute gradients of a loss function and send updates to a master process.  An alternate algorithm, Elastic Averaging SGD~\cite{easgd}, is also available.  

In Downpour SGD, one process is assigned to be the master and the others are assigned to be workers.  The master and each worker have a copy of the model to be trained. Each worker has access to a subset of the training data.

During training, a worker reads one batch of training data and computes the gradient of the loss function on that batch. The worker sends the gradient to the master, which uses it to update its model weights. The master sends the updated model weights to the worker, which then repeats the process with the next batch of training data.

\begin{figure}
  \caption{Diagram illustrating the Downpour SGD training algorithm.}
  \label{fig:downpour}
  \centering
    \includegraphics[width=0.5\textwidth]{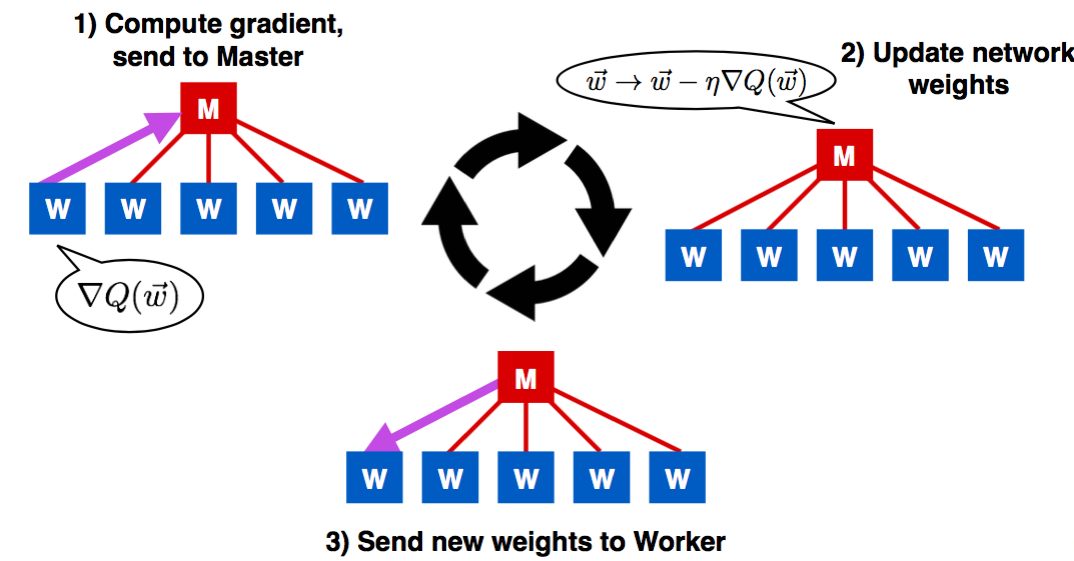}
\end{figure}

See Fig.~\ref{fig:downpour} for an illustration of the Downpour SGD training procedure.  

In the Elastic Averaging SGD algorithm, worker processes are connected to a master via an elastic force that periodically `pulls' the weights closer to one another.  Workers train independently and communicate with the master only via the elastic force, allowing each worker to explore a different region of the model parameter space.

Training proceeds asynchronously by default, with worker processes exchanging weight information with the master one by one~\cite{downpour}.  Synchronous training is also available; in this case the master processes weights from all workers simultaneously. In addition to the canonical training configuration with one master process and several workers, the \textit{mpi\_learn} framework also supports a hierarchical configuration in which there are several master processes, each coordinating a group of workers and reporting to a higher-level master.  

\subsection{User Interface}
The user interface to the \textit{mpi\_learn} code consists of three main components, each handled via a Python class:
\begin{itemize}
\item The training procedure is specified via an Algo class that stores information such as the batch size, choice of optimization algorithm, loss function, and any tunable training parameters such as the learning rate.
\item The DNN model is specified via a ModelBuilder class that provides instructions for constructing a Keras model.  The model architecture can be read from a JSON file or specified via Keras code.  Using the Tensorflow backend to Keras, it is possible to achieve model parallelism by specifying a device (GPU or CPU) for each layer of the model individually.
\item Input data is specified via a Data class that provides a data generator for use during the training phase. The user may provide a list of input file paths, which are divided evenly among all worker processes during training.
\end{itemize}
More details on the code can be found on the \textit{mpi\_learn} Github page \cite{github}.

\section{Experiment}

The training time speedup for a benchmark neural network model was evaluated on two systems:
\begin{itemize}
\item A Supermicro server with 28 cores and eight NVidia GTX1080 GPUs. 
Communication between processes is accomplished via shared memory, as all processes are on the same node.
\item The ALCF Cooley \cite{cooley} GPU cluster, with 126 nodes, each having 16 cores and 1 NVidia K80 GPGPU. Nodes are interconnected with FDR Infiniband.
\end{itemize}
The performance results reported in this paper are in no way a comparison of the systems detailed above; they simply demonstrate the speedup of the training procedure when \textit{mpi\_learn} is used to distribute the training over multiple GPUs. Further performance improvement could be obtained by tuning the software to the specific architecture of the system used.

The \textit{mpi\_learn} framework was used to train a recurrent neural network to classify simulated collision events from high-energy particle detectors at the CERN Large Hadron Collider~\cite{lhc}. The model consists of an LSTM network~\cite{lstm} with 20 hidden units, followed by a softmax output over three different categories of collision events.
The dataset was created using the Delphes simulation framework~\cite{delphes}.
The input data consists of 100 files of 9500 samples each, totaling 50GB.
This model takes several hours to train on a node with a single GPU.

\begin{figure}
  \caption{Model accuracy after 10 training epochs as a function of the number of workers used. The model performance slowly decreases at high worker counts because of workers training on outdated model information.}
  \label{fig:accuracy}
  \centering
    \includegraphics[width=0.5\textwidth]{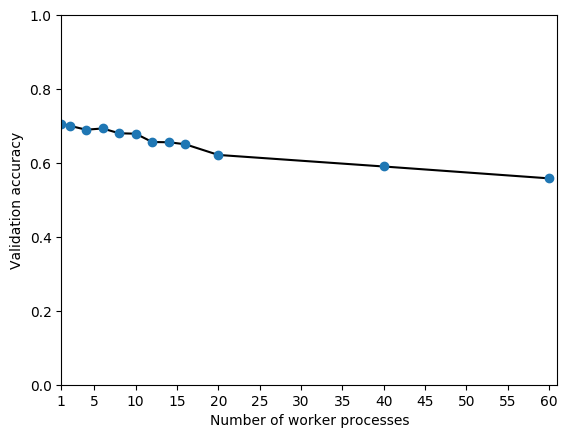}

\end{figure}

The purpose of this paper is not to evaluate the performance of the model~\cite{dannypaper} but rather to evaluate how much faster this model can be trained when multiple GPUs are utilized.  
As shown in figure \ref{fig:accuracy}, the model accuracy slightly degrades with increased number of workers.  This occurs because of the so-called stale gradient issue: workers training on outdated model parameters produce suboptimal gradient updates.  The issue can be mitigated by a suitable choice of SGD momentum~\cite{omnivore}.

\section{Performance}

The model is trained several times with various numbers of worker processes, using a batch size of 100 samples.  
The data in the training set is divided evenly among all workers.  Training continues until each worker has processed
its training data a fixed number of times (ten, in this case).  For each batch of training data, a worker must compute the gradient 
of the loss function, send the gradient to the master process, and receive updated weights from the master after it applies the gradient
update.

Validation of the model's accuracy is performed by the master process using a held-out test set.  Validation can be a bottleneck in
the training process because it is performed serially; the frequency of validation can be adjusted as needed to minimize its impact
on the total training time.  

\begin{figure}
  \caption{Speedup performance on the Supermicro server with 8 GPUs, as a function of the number of workers used for training, with a batch size of 100 samples. The red dotted diagonal indicates 1:1 speedup.}
  \label{fig:speedupALCF}
  \centering
    \includegraphics[width=0.5\textwidth]{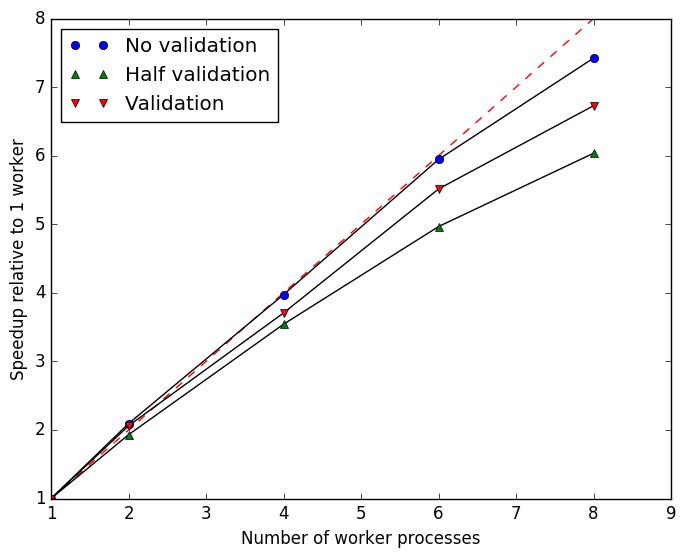}
\end{figure}

\begin{figure}
  \caption{Training speedup for the benchmark model on the ALCF Cooley cluster with 1 GPU per node, as a function of the number of workers used for training, using a batch size of 100 samples. The red dotted diagonal indicates 1:1 speedup.}
  \label{fig:speedupSM}
  \centering
    \includegraphics[width=0.5\textwidth]{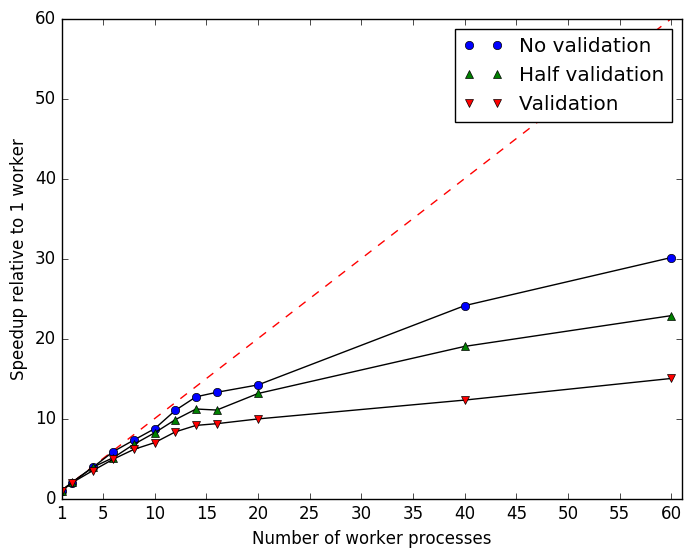}
\end{figure}

For each training run, the speedup is computed with respect to the time taken by \textit{mpi\_learn} using a single worker process.
Results are shown in figures \ref{fig:speedupALCF} and \ref{fig:speedupSM}.

The time needed to train the model with \textit{mpi\_learn} and a single worker process is also
compared to the training time obtained using Keras alone. 
The times are similar, indicating that the training overhead from the \textit{mpi\_learn} framework itself is small.  

For up to 10 worker processes, the speedup is roughly linear with the number of workers.  This indicates that the training framework
can fully exploit the resources of a multi-GPU node such as the Supermicro server used here.

The speedup deviates from linearity with increasing number of workers. For 60 worker nodes, we observe a speedup of 30 with respect to 
the nominal training time for this choice of batch size.  The deviation from linearity is driven by the time needed for the master process to update
the weights of the network and transmit them back to the workers. Because the frequency of weight updates is inversely proportional to 
the batch size, increasing the batch size can alleviate this bottleneck and speed up the training procedure, as shown in
Table~\ref{table:batches} for the example of 20 worker processes.  

The higher the amount of validation the earlier the linear scaling will break, because the constant amount of time spend in validation that cannot be compressed by adding more workers to the training part.
This is confirmed with the trend of getting better speedup when decreasing the amount of validation.

\begin{table}[ht]
\caption{Speedup obtained with various batch sizes, with respect to a batch size of 100, with 20 workers training the model.}
  \centering
\label{table:batches}
\begin{tabular}{|c|c|}
	\hline
	Batch Size & Speedup\\
	\hline
    10 & 0.1 \\
    100 & 1.0 \\
    500 & 3.0 \\
    1000 & 4.1 \\
    \hline
\end{tabular}

\end{table}

\section{Discussion}

The \textit{mpi\_learn} package provides a convenient interface for training Keras models in a distributed fashion using the MPI protocol.  
The system is straightforward to use with most models and can be flexibly customized.  

Performance has been evaluated on up to eight GPUs on a single server and on up to 60 GPUs on the ALCF Cooley cluster.  
The results demonstrate a linear speedup with the number of workers in a certain regime, and in particular allows full usage of the resources of multi-gpu servers.
By providing this training framework, we hope to make it easier for researchers in the sciences and other fields to fully harness available computing resources and benefit from existing distributed training algorithms.  
The framework is lightweight enough to be used without extensive configuration.  
It can be used on any MPI-enabled machine or cluster, making it especially practical for training using supercomputing resources. 
These properties facilitate quick prototyping of large deep neural models and training using many GPUs and/or CPUs, an ability that will become more important as deep learning continues to spread to new application areas.  

\section{Acknowledgments}
This research used resources of the Argonne Leadership Computing Facility, which is a DOE Office of Science User Facility supported under Contract DE-AC02-06CH11357. We are grateful to  Venkatram Vishwanath and Andrew Cherry for their support with Cooley. We acknowledge NVIDIA, SuperMico  and the Kavli Foundation for their support of "\textit{iBanks}", the Caltech HEP AI Tower. This work is partially supported  by the United States Department of Energy, Office of High Energy Physics under Contract No. DE-SC0011925 and DOE OHEP Research Technology, Computational HEP and Fermilab under Contract No. DE-AC02-07CH11359.

\addtolength{\textheight}{-12cm}   


\end{document}